\documentclass[article,onecolumn]{mn2e}
\usepackage{times}
\input{psfig.sty}
\newif\ifAMStwofonts

\newcommand{\e}{\epsilon}

\begin{document}  

\title[Photospheric signatures imprinted on the GRB
spectra]{Photospheric signatures imprinted on the $\gamma$-ray burst
spectra}

\author[Ramirez-Ruiz]{Enrico Ramirez-Ruiz$^{1}$ \\ School of Natural
Sciences, Institute for Advanced Study, Einstein Drive, Princeton, NJ
08540, USA.\\ $^{1}$Chandra Fellow.}

\date{}

\maketitle

\label{firstpage}

\begin{abstract}
A solution is presented for the spectrum of high-energy GRB photons
confined to a quasi-thermal baryonic photosphere. The solution is
valid in the steady-state limit assuming the region under
consideration is optically thick to the continuously injected
photons. It is shown that for a high luminosity photosphere, the
non-thermal electrons resulting from $\gamma$-ray Compton cooling lose
their energy by upscattering the soft thermalised radiation. The
resulting spectral modifications offer the possibility of diagnosing
not only the burst comoving luminosity but also the baryon load of the
ejecta. This model leads to a simple physical interpretation of X-ray
rich bursts and anomalous low-energy slopes.
\end{abstract}


\section{Introduction}
A quasi-thermal photosphere is expected in all GRB ejecta (Cavallo \&
Rees 1978; Goodman 1986; Shemi \& Piran 1990), either due to pairs or
to baryonic electrons, in addition to a possible non-thermal component
from dissipation (shocks or reconnection). Preponderance of one or the
other depends on whether the photosphere occurs inside or outside the
saturation radius where the bulk Lorentz factor saturates to the
dimensionless entropy of the outflow. Below the saturation radius, the
photospheric luminosity dominates whereas above it the greater part of
the energy is in kinetic form. In the latter case, the energy
available in the from of radiation or pairs depends on the variable
portion of the outflow and the efficiency of dissipation in shocks
(Daigne \& Mochkovitch 2002; Ramirez-Ruiz et al. 2002), or on the
reconnection efficiency (M\'esz\'aros \& Rees 2000).

In this Letter we investigate the relationship between the
quasi-thermal baryon-related photosphere in relativistic outflows, and
the internal shock arising around this limiting radius.  This
photosphere is a source of soft thermal radiation, which may be
observationally detectable in some GRB spectra (Murakami et
al. 1991). In these models, dissipation happens whenever internal
shocks develop in the ejecta (Rees \& M\'esz\'aros 1994), which
reconverts some fraction of the kinetic energy into radiation. If this
dissipation takes place below the photosphere, the non-thermal shock
luminosity could compensate for the adiabatic decrease of the
post-saturation photospheric luminosity, so that the radiative
efficiency in the outflow would be high. The relative roles of
photosphere and shocks are reexamined in Section 2. A solution is
presented in Section 3 for the spectrum of high energy $\gamma$-rays
confined to a quasi-thermal baryonic photosphere and the possible non
thermal spectral distortions in it. The relation of theses ideas to
GRB phenomenology is outlined in Section 4.

\section{Photospheres and Shocks}
Consider a relativistic wind outflow expanding from some initial
minimum radius $r_0=c \delta t =10^7 r_{0,7}$ cm, where the wind
baryon load $\dot{M}$ is parametrized by a dimensionless entropy
$\eta=L_0/\dot{M}c^2$.  We then assume that the actual value of $\eta$
(or $L_0$) is unsteady.  The Lorentz factor saturates to $\Gamma \sim
\eta$ at a saturation radius $r_\eta/r_0 \sim \eta$ where the wind
energy density, in radiation or in magnetic energy, drops below the
baryon rest mass density in the comoving frame (Shemi \& Piran 1990).
The location of the baryonic photosphere, where $\tau_T=1$, due to
electrons associated with baryons, is
\begin{equation}
 r_{\tau}={{\dot M} \sigma_T \over 4\pi m_p c \Gamma^2} \approx
 10^{13} L_{0,52}\eta_2^{-3} ~{\rm cm},
\label{eqn:photo}
\end{equation}
where $\eta_2=\eta/10^2$ and $L_{0,52}=L_0 /10^{52}\; {\rm erg}$. The
above equation holds provided that $\eta$ is low enough that the wind
has already reached its {\it terminal} Lorentz factor at
$r_{\tau}$. This requires (M\'esz\'aros \& Rees 2000)
\begin{equation}
\eta \le \eta_\ast =\left({L_0 \sigma_T \over 4\pi m_p c^3
r_0}\right)^{1/4} \simeq 10^3 L_{0,52}^{1/4} r_{0,7}^{-1/4}.
\end{equation}

If the value of $\eta$ at the base increases by a factor $\ge 2$ over
a timescale $\delta t$, then the later ejecta will catch up and
dissipate a significant fraction of their energy at some radius
$r_{\iota} > r_\eta$ given by
\begin{equation}
r_{\iota} \sim c \delta t \eta^2 \sim 3 \times 10^{14} \delta t_{0}
\eta_2^2~{\rm cm}.
\label{eq:rint}
\end{equation}
The shocks cannot occur when $r < r_\eta$ since in this region both
shells accelerate at the same rate $\Gamma \propto r$ and do not catch
up. In order for a shock to occur above a photosphere which is in the
coasting region
\begin{equation}
\eta \ge \eta_\tau \approx 50 L_{0,52}^{1/5} \delta t_{0}^{-1/5},
\end{equation}
if one takes $r_0\sim c\delta t$, where $\delta t_0=\delta t/1\;{\rm
sec}$. If, on the other hand, $\delta t\sim 10^{-3}$ sec, then
$\eta_\tau \sim 200$.

The initial wind starts to decelerate when it has swept up $\sim
\eta^{-1}$ of its initial mass. For sufficiently high $\eta$, the
deceleration radius can formally become smaller than the collisional
radius of equation (\ref{eq:rint}). This requires
\begin{equation}
\eta \ge \eta_{\rm d} \approx 8 \times 10^2 L_{0,52}^{1/8}
t_{w,1}^{1/8} n_{0}^{-1/8} \delta t_{0}^{-3/8}.
\label{eq:etab}
\end{equation}
This deceleration allows slower ejecta to catch up, replenishing and
re-energising the reverse shock and boosting the momentum in the blast
wave. 

\begin{figure*}
\centerline{\psfig{figure=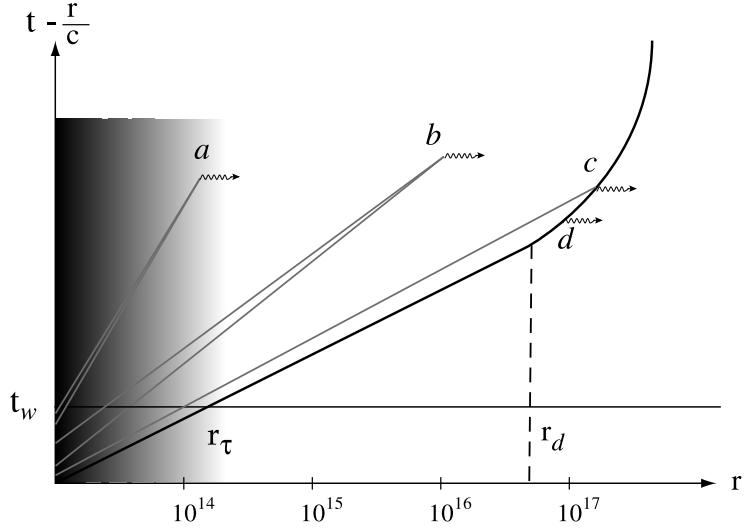,angle=0,width=0.55\textwidth}}
\caption{
Schematic spacetime diagram in source frame coordinates of a
relativistic outflow. The axes (logarithmic) are {\it r} versus $t-
(r/c)$, where $t$ is time measured by a distant observer, and is zero
when the burst is observed to start. In this plot, light rays are
horizontal lines. The primary gamma-ray emission is assumed to
continue, with a quasi-steady luminosity $L_w$, for a time $t_w$. If
$\Gamma$ fluctuates by a factor of $\sim$ 2 around its mean value,
relative motions within the outflowing material give rise to internal
shocks.  Decreasing $\eta$ values lead to world lines further to the
left. In case (a), $\eta < \eta_\tau$ and the dissipation occurs when
the wind is optically thick. In case (b), with $\eta_{\rm d}> \eta >
\eta_\tau$, ejecta collide in an optically thin region before reaching
the contact discontinuity. The contact discontinuity and the forward
shock are being decelerated because of the increasing amount of
external matter being swept up (giving rise to a long-term afterglow;
case [d]), so that they lag behind the light cone by an increasing
amount $\Delta r$ , whose increase with r is steeper than linear. This
deceleration allows ejecta to catch up and pass through a reverse
shock just inside the contact discontinuity (case [c]).  }
\label{fig:diagram}
\end{figure*}

Fig. 1 shows the schematic world-lines of a relativistic outflow with
a range of Lorentz factors. We identify three types of contributions
to the observed time history, each with a different character. For a
relatively low Lorentz factor $\eta < \eta_\tau$, as in curve (a) of
Fig. 1, the radius $r_\iota < r_\tau$, and the dissipation occurs
before the wind is optically thin. Below the baryonic photosphere,
shocks would occur at high optical depths. Section 3 investigates the
relationship between the quasi-thermal baryon-related photosphere in
relativistic outflows and the internal shocks arising near this
limiting region. For a larger Lorentz factor $\eta > \eta_{\rm d}$,
corresponding to curve (c) of Fig. 1, the ejecta would expand freely
until the contact discontinuity had been decelerated by sweeping up
external material. It would then crash into the reverse shock,
thermalising its energy and boosting the power of the afterglow. The
impact of such collisions on the prompt emission have been studied in
Ramirez-Ruiz, Merloni \& Rees (2001). In curve (b), with intermediate
$\eta$, deceleration occurs at radii $r_{\rm d}>\;r_\iota$, and
dissipation takes place when the wind is optically thin (i.e. when it
is most effective).

\subsection{Photospheric and Internal Shock Luminosity}
The lab-frame baryonic photospheric luminosity $L_{\tau}$ and
dimensionless temperature $\Theta_{\tau}$ evolve as $(L_{\tau}/L_0) =
(\Theta_{\tau}/\Theta_0) = (r_{\tau}/ r_\eta)^{-2/3}$, where
$\Theta_0=kT_0/m_ec^2 \simeq 2 L_{0,52}^{1/4}r_{0,7}^{-1/2}$ is the
initial temperature at $r_0$. Using the expression for $r_\tau$, this
can be written as $(L_{\tau}/L_0) = (\eta/\eta_\ast)^{8/3}$ for
$\eta<\eta_\ast$ and $1$ otherwise.

The internal shocks in the wind can dissipate a fraction of the
terminal kinetic energy luminosity $L_0$ above the saturation radius
$r_\eta$, $L_\iota = \e_\e \e_\iota L_0 \sim
10^{-1}(3\e_\e)(3\e_\iota)L_0~$, where $\e_\e \e_\iota$ is a
bolometric radiative efficiency when the cooling timescale is shorter
than the dynamical time. If the Poynting flux provides a fraction
$\alpha$ of the total luminosity $L_0$ at the base of the wind (at
$r_0$), the magnetic field there is $ B_0 \sim
10^{10}\alpha^{1/2}L_{0,51}^{1/2}\delta t_{0}^{-1}$ G. The
comoving magnetic field at $r_\iota$ is $B_\iota=B_0 (r_0/r_\eta)^2
(r_\eta/r_\iota) \sim 10^4\alpha^{1/2}L_{w,51}^{1/2} \delta t_{0}^{-1}
\eta_2^{-3}$ G. If the electrons are accelerated in the
dissipation shocks to a Lorentz factor $\gamma=10^3 \gamma_3$, the
ratio of the synchrotron cooling time to the dynamic expansion time in
the comoving frame is $(t_{\rm sy}/t_{\rm adi})_\iota \sim 5\times
10^{-3}\alpha^{-1}L_{0,51}^{-1}\gamma_3^{-1} \delta t_{0} \eta_2^5$ so
a very high radiative efficiency is ensured even for $\delta t$ as
high as seconds. It is therefore clear that a magnetic field can
ensure efficient cooling even if it is not strong enough to be
dynamically significant (i.e. even for $\alpha \ll 1$).

For a given $L_0$, an individual burst is characterized by an average
$\eta$. For $\eta>\eta_\ast$, the photosphere arises in the
accelerating region, and in this region shocks are not possible.  For
$\eta_\tau <\eta < \eta_\ast$, we have $L_{\tau}> L_\iota$, so that
the baryonic photospheric component dominates the non-thermal internal
shock component in a bolometric sense. Low values of $\eta$ lead to
further out, weaker photospheres and, at the same time, relative
stronger shocks occurring near the photosphere. For $L_{\tau}\leq
L_\iota$, the thermal photospheric peak will tend to blend with the
synchrotron peak, resembling the canonical non thermal GRB
spectrum. For lower $\eta < \eta_\tau$, the thermal peak is even
softer with $L_{\tau} \ll L_\iota$, while shocks occur closer in and
produce harder synchrotron peaks (Ramirez-Ruiz \& Lloyd-Ronning
2002). The corresponding shocks occur below (or near) the baryonic
photosphere where the scattering depth of the shells is larger
(Panaitescu et~al. 1999; Ramirez-Ruiz \& Lloyd-Ronning 2002, Kobayashi
et~al. 2002). Shocks which occur inside the photosphere may also
induce Alfv\'en waves and these waves can be efficiently damped for
$r<r_\tau$ (i.e. $\tau_T>1$).

\section{The effects of a Reprocessing Photosphere on the GRB spectra}
We consider here the problem of Compton downscattering of X-rays and
$\gamma$-rays confined to an optically thick baryonic photosphere.
The electrons are assumed to have zero temperature, which is
appropriate as long as the photon energies are much larger than
$\Theta_\tau$. For simplicity we consider here only shocks producing a
comoving spectrum with $h\nu' \leq 511$ keV so that photon-photon pair
production is negligible. In this case, the pair photosphere is not
sufficient to alter significantly the spectrum of the baryonic
photosphere. In the following we shall assume the acceleration of
electrons in the shocks to be impulsive, and therefore it has to take
place in a very limited volume of the interacting shell (i.e the
emitting particles do not provide an additional heating source).

We assume that GRB photons are continuously and uniformly injected at
a rate $\dot{\phi}_{i}(\e)$, where $\e$ is the dimensionless photon
energy in units of $m_e c^2$. For future use we take a power-law
photon injection of the form $\dot{\phi}_{i}(\e)= \Psi \e^{-\beta}$,
for $\e \leq \e_b$. Under the assumption that photosphere is optically
thick, a steady photon density distribution $\phi (\e)$ can be found
after specifying the photon production rate and the electron density
$n_e$ of the region under consideration. This distribution is the
solution to a kinetic equation where the main energy-loss mechanism is
Compton scattering. $\dot{\phi}_{i}(\e)$ satisfies an equation of the
form (Arons 1971)
\begin{equation}
0=\dot{\phi}_{i}(\e)+\int_{\e}^{\e_b} d\e' \phi(\e')\sigma (\e',\e)
-\phi(\e)\int_{\e/(1+2\e)}^{\e}d\e'\sigma (\e',\e),
\label{eqn:kineq}
\end{equation}
where the three terms correspond to the injection of primary photons,
Compton downscattering to $\e$ from higher energies, and Compton
scattering out of the energy $\e$ to lower energies,
respectively. $\sigma (\e',\e)$ in equation (\ref{eqn:kineq})
represents the probability per unit time that a photon with energy
$\epsilon$ will Compton scatter from $\e'$ to $\e$ (Jauch \& Rohrlich
1980)
\begin{equation}
\sigma (\e',\e)={3 \over 8} n_e c \sigma_T \left({1 \over
{\e'}}\right)^2 \left[ {\e' \over \e} + {\e \over \e'} - 2\left({1
\over \e}-{1 \over \e'}\right) + \left({1 \over \e}-{1 \over
\e'}\right)^2\right]\;.
\label{eqn:scprob}
\end{equation}
 
\begin{figure*}
\centerline{ \psfig{figure=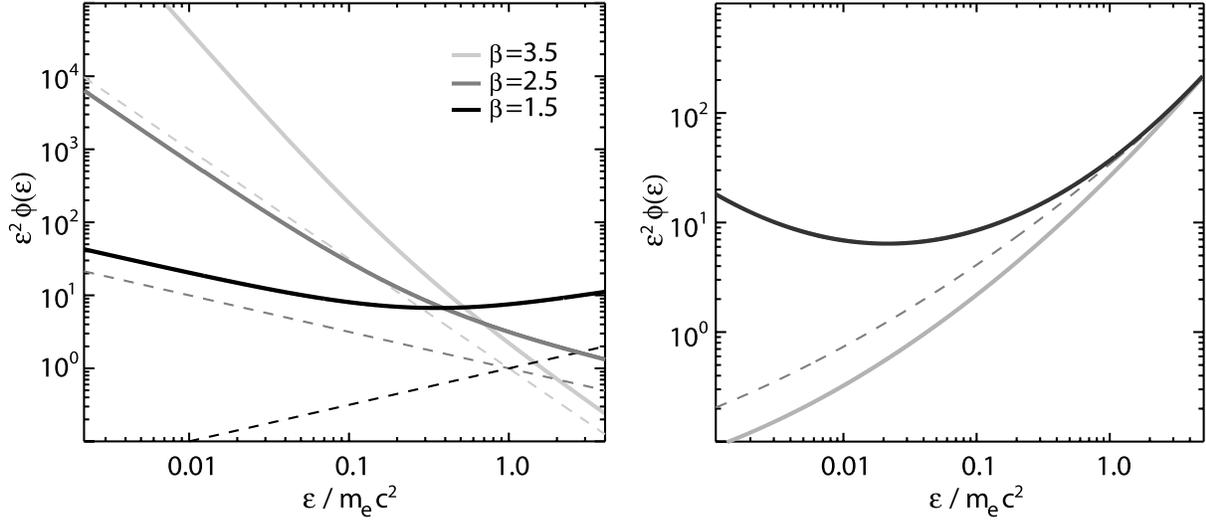,angle=0,width=0.9\textwidth}}
\caption{The photon density distribution from a power-law injection of
$\gamma$-rays. The normalizations of $\phi(\e)$ correspond to
$\Psi/(n_e\sigma_Tc)$=1. {\it Left Panel:} Power-law injection of
primary photons with index $\beta$ and $\e_b$=5 (solid lines). The
dotted curves correspond to the density of unscattered photons. {\it
Right Panel:} The evolution of the photon density distribution in the
presence of a power-law injection of primary photons with index
$\beta=1.5$ and $\e_b$=10. The black solid line gives the total photon
density distribution. The grey solid line corresponds to the density
of unscattered photons, and the grey dashed curve to the sum of
unscattered and singly scattered photons.}
\label{comptfig}
\end{figure*}

The integral equation (\ref{eqn:kineq}) with the scattering rate
(\ref{eqn:scprob}) can be solved numerically by iteration for any
$\e$, where the iterative steps correspond to subsequent orders of
Compton scattering. In Fig. \ref{comptfig} are plotted representative
spectra for a power-law injection of high-energy photons in the
conventional coordinates $\epsilon$ and
$\epsilon^2\phi(\epsilon)$. The dotted curves give the density of the
unscattered photons, $\phi_u$, produced at the power law rate. The
solid curves, on the other hand, give the total photon density which
includes both unscattered and scattered photons.  Here $\phi_u$ is
given by $\phi_u (\e) = \dot{\phi}_{i}(\e)/[n_e \sigma(\e)c]$. The
calculations shown in Fig.  \ref{comptfig} approximate the electrons
in the medium to have zero temperature and assume that Compton
scattering is the main interaction process for $\gamma$-rays.  The
results shown in Fig. \ref{comptall} expand this calculation to
include the role of secondary electrons (produced by Compton
downscattering) under the assumption that they lose their energy
mainly by upscattering the soft thermalised radiation (see Section 3.1).

The right panel of Fig. \ref{comptfig} shows the evolution of
the photon density distribution for subsequent orders of Compton
scattering. The black solid curve gives the total photon density
distribution which includes both unscattered and scattered
photons. The grey solid line corresponds to the density of unscattered
photons, and the grey dashed curve to the sum of unscattered and
singly scattered photons. The density of photons scattered once and
more is therefore given by the difference between the black solid and
grey dashed curves.

When $\e \ll 1$, it is straightforward to present a simplified
discussion of the problem. Analytical solutions are derived here
largely to illustrate what may not be obvious from the numerical
treatment. The last integral in equation (\ref{eqn:kineq}) can be
rewritten as
\begin{equation}
\int_{\e/(1+2\e)}^{\e}d\e'\sigma (\e',\e)= n_e c \sigma (\e),
\end{equation}
where $\sigma(\e)$ is the Klein-Nishina cross section. In the case of
$\e \ll 1$, $\sigma(\e) \approx \sigma_T$. The Fokker-Plank equation
can then be derived by expanding (\ref{eqn:kineq}) in powers of the
small quantity $\e-\e'$ (Ross et al. 1978)
\begin{equation}
\dot{\phi}_{i}= -n_e\sigma_Tc {d \over d\e}\left[{7\over 10}\e^6{d
\over d\e}\left({\phi \over \e^2}\right)+ \e^2\phi \right]
\label{fpe}
\end{equation}
where the first term describes the photon energy dispersion, whereas
the second characterizes the systematic changes in $\e$. Equation
(\ref{fpe}) differs from the equation of Kompaneets (1957) in that
there is an additional dispersion term. For a power-law injection, the
solution to equation (\ref{fpe}) reads
\begin{equation}
\phi(\e)\approx{\Psi \e^{-2} \over n_e \sigma_T c (\beta -1)}
\left[\e^{1-\beta}\left(1 + {7\e \{\beta +3\} \over
10}\right)-\e_b^{1-\beta}\left(1 + {14 \over 5}\e \right)\right].
\label{sol}
\end{equation}
Here we have kept only the first two powers of $\epsilon$. Provided
that $\e_b \geq 1$ and $\beta>1$, from equation (\ref{sol}) it follows
that $\phi / \dot{\phi}_{i} \propto \epsilon^{-1}$.  As illustrated in
Fig. \ref{comptfig}, most of the luminosity at low energies results
from $\gamma$-ray downscattering.

\subsection{The role and fate of the secondary, energetic electrons} 
So far we have approximated the electrons in the medium to have zero
temperature and have assumed that Compton scattering is the main
interaction process for $\gamma$-rays.  Compton cooling of
$\gamma$-rays, however, produces a population of secondary energetic
electrons. The fate of these electrons depends on the composition,
ionization state and magnetic field present in the medium, as well as
on the energy density of the background radiation. When the energy
density in the (black body) photospheric photons is much larger than
the energy in scattered photons (i.e. $L_{\tau}> L_\iota$), the
secondary electrons will lose energy mostly in collisions with the
photospheric photons and form a secondary photon spectrum.  This is
the situation considered in this section.

To first approximation, the scattering medium can be considered as a
cold plasma if the time between successive scatterings, $t_\tau \sim
\epsilon m_e c /(U_\iota \sigma_T)$, is much longer than the
characteristic cooling time of the secondary electrons, $t_\gamma \sim
{3 \over 4} m_e c/(\gamma U_\tau \sigma_T)$. Here $U_\iota$ and
$U_\tau$ are the energy densities of the continuously injected photons
and the photospheric radiation, respectively. The secondary electrons
created by downscattering Compton cooled before the next generation of
hot electrons is created if $U_\tau > {3 \over 4} U_\iota/ (\epsilon
\gamma$). Therefore, the bulk of the injected radiation is likely to
be scattered by cooled particles when $L_{\tau}> L_\iota$. In what
follows we assume the scattering medium to be ``cold'' since it
greatly reduces the calculations. A detailed discussion of the
relevance of a hot scattering medium, its steady state solutions for
various input spectra, and its range of applicability (i.e $L_{\tau}
\leq L_\iota$) will be presented in future work.

The injected photons repeatedly downscatter to form a steady state
distribution. During this process, a population of secondary,
energetic electrons is injected. Down-scattering of $\gamma$-rays on a
cold medium produces secondary electrons at a rate
\begin{equation}
\dot{n}_\gamma(\gamma)=\int_{\e_m}^{\e_b} d\e\phi(\e)\sigma(\e,\e-\gamma+1),
\label{elec}
\end{equation}
where $\gamma$ is the electron Lorentz factor and $\phi$ is the steady
state photon density distribution. Here $\e_m={1\over
2}[\gamma-1+\{(\gamma-1)^2+2(\gamma-1)\}^{1/2}]$ is the minimum energy
required for a photon to up-scatter an electron from rest to the
energy $\gamma$.

The maximum energy of the secondary electrons is then given by
$\gamma_b=1+ 2\e_b^2/(1+2\e_b)$. These secondary electrons could lose
their energy mainly through synchrotron or Compton energy-loss
mechanisms. One or both of these mechanisms may be present, depending
on the bulk Lorentz factor, the isotropic equivalent total energy of
the burst, and the efficiency of dissipation in shocks. In the
previous section we discussed the criteria for the non-thermal
component to dominate over, or be dominated by, the photospheric
thermal component.  A photosphere should be prominent in the lowest
baryon load cases.The black body photons can then act as seeds for
scattering to higher energies, if there is substantial amount of
energy in scattering centers. Alfv\'en waves generated by magnetic
field reconnection or MHD turbulence can act as such centers. Repeated
scattering on the Alfv\'en waves acts in the same way as
Comptonisation off hot electrons.

Electrons resulting from down-scattering will therefore lose energy
mostly in collisions with the photospheric emission when the energy
density in the photospheric blackbody exceeds the scattered one. The
steady state electron distribution (Blumenthal \& Gould 1970) is given
by
\begin{equation}
n_\gamma(\gamma)= {\int_\gamma^{\gamma_b}
d\gamma'\dot{n}_\gamma(\gamma') \over {4 \over 3}\gamma^2\sigma_Tc
\int_0^{\infty}d\e_\Theta \e_\Theta \phi_\Theta (\e_\Theta)},
\end{equation}
where $\phi_\Theta (\e_\Theta)$ is the blackbody photon density
distribution.

The production rate of secondary high-energy photons
(i.e. $\Theta>\Theta_\tau$) produced by the upscatter of blackbody
photons by the secondary electrons is given by
\begin{equation}
\dot{\phi}_s(\e)=\int_{1+\e}^{\gamma_b} d\gamma n_\gamma (\gamma)
\sigma_\Theta (\gamma, \gamma-\e),
\end{equation}
where $\sigma_\Theta (\gamma, \gamma')$ is the rate of scatterings of
electrons from $\gamma$ to $\gamma'$ on blackbody photons. Zdziarski
(1988) has given an explicit expression for $\sigma_\Theta (\gamma,
\gamma')$. The assumption of negligible photon-photon pair production
guarantees that collisions between secondary electrons and black body
photons take place in the $\epsilon \ll 1$ limit. The total rate of
photon injection is then given by $\dot{\phi}_s(\e) +
\dot{\phi}_i(\e)$. This photon rate results in a new steady state
photon distribution (\ref{eqn:kineq}), and, in turn, a new secondary
electron production rate (\ref{elec}). The production rate of
secondary photons resulting from scatterings of the new secondary
energetic electrons with the black body photons is calculated and
subsequently added to the total rate of photon injection. This
procedure is repeated until convergence is achieved
(i.e. $\dot{\phi}_s^{n}+ \dot{\phi}_i^{n}$ for $n=1,2,3,...$). For
each step the integral equation (\ref{eqn:kineq}) is solved with the
scattering probability (\ref{eqn:scprob}) numerically by iteration
(see e.g. right panel of Fig. \ref{comptfig}).

\begin{figure*}
\centerline{ \psfig{figure=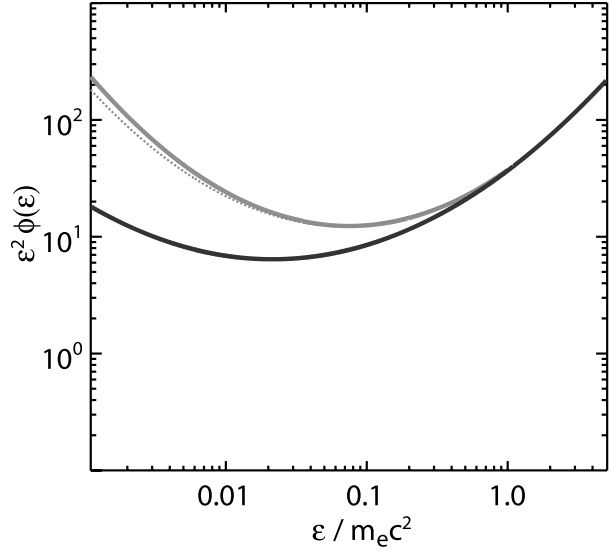,angle=0,width=0.45\textwidth}}
\caption{Illustration of the effect of Compton scattering of
photospheric blackbody photons at $T_\tau=$0.5 keV by secondary
electrons.  The role of secondary electrons is included under the
assumption that they lose their energy mainly by upscattering the soft
thermalised radiation and that they Compton cooled before the next
generation of hot electrons is created. The production rate of
secondary photons resulting from scatterings of the secondary
energetic electrons with the black body photons is calculated and
subsequently added to the total rate of photon injection. This
procedure is repeated until convergernce is achieved. The dark solid
line gives the spectra without the effects of blackbody up-scattering
(Fig. \ref{comptfig}).  The grey solid line, on the other hand,
illustrates the spectral distortions arising from the inclusion of
photospheric photons while the dotted line gives the photon
distribution after only the first iteration procedure.  The primary
photons are injected with a power-law distribution with $\beta=1.5$
and $\e_b=10$. The normalization of $\phi(\e)$ correspond to
$\Psi/(n_e\sigma_Tc)$=1.}
\label{comptall}
\end{figure*}

Fig. \ref{comptall} illustrates the effect of Compton up-scattering of
photospheric blackbody photons at $\Theta_\tau \sim 10^{-3} m_ec^2$ by
secondary electrons. The dotted curve shows the steady state photon
distribution with the inclusion of the soft black-body photons while
the black dashed line gives the photon distribution after only the
first iteration procedure.  The photons from photospheric upscattering
dominate the spectrum at $\e \ll 1$. The highest energy (upscattered)
photospheric photons are produced by relativistic electrons from
Compton downscattering of the primary photons. At lower energies, on
the other hand, the upscattered photospheric photons are due to
electrons arising from multiple scattered photons. The importance of
the upscattered photospheric photons at $\epsilon \leq 0.1$ depends
mainly on the low-energy tail distribution of the primary photons and
thus its effect on the photon spectra increases with subsequent
iterations.  This emission could be responsible for the X-ray excess
observed in some GRBs.  In contrast, at higher energies, the injected
photon distribution (i.e. $\dot{\phi}_s^{n} + \dot{\phi}_i^{n}$) is
not significantly alter by subsequent iterations and the role of
upscattered photospheric photons to the high-energy spectra can be
estimated with only a few iterations.

One does not  expect this scattered photospheric component to be a pure
Wien spectrum, since in order to change the shape of a soft spectrum
and make it into a dilute black body spectrum would require that each
photon be scattered $m_ec^2/h\nu$ times by a population of electrons
in thermal equilibrium. This requires an optical depth much larger
than unity, which is generally not the case anywhere near the
photosphere. The scattering depth per shock due to pairs is also
unlikely to be much larger, because the scattering and the
pair-formation cross sections are comparable, and unless dissipation
and pair formation occurs uniformly throughout the entire volume,
down-scattering of photons above the pair threshold rapidly leads to
self-shielding (M\'esz\'aros et al. 2002).

\section{Discussion}
The standard internal shock model of GRB is generally assumed to
produce its observed non thermal radiation by synchrotron (or possibly
inverse Compton) process. Here, in addition to synchrotron we have
also considered in more detail the role of the outflow photosphere and
of possible non thermal distortions in it. A strong baryonic
photosphere component should be present at the beginning of
bursts. However, the farther beyond the coasting radius the
photosphere occurs, the weaker its energy fraction is relative to the
primary injected photons arising from internal shocks, because its
energy drops as $r^{-3/2}$.

The problem of reprocessing X-rays and $\gamma$-rays by Compton
scattering could be the key to understanding the X-ray excess above
the power-law extrapolation from higher energies observed in a
non-negligible fraction of bursts (Kippen et al. 2001; Heise et
al. 2001; Amati et al. 2002; Lloyd-Ronning \& Ramirez-Ruiz 2002;
Sakamoto et al. 2005), although smoothing and softening could be
stronger where there is substantial pair formation (Pilla \& Loeb
1998; Ghisellini \& Celotti 1999; M\'esz\'aros et al. 2002; Pe'er \&
Waxman 2004). If there were a power-law non-thermal spectrum in which
significant fraction of the resulting photons are above 1 MeV in the
comoving frame, then pair production will change the
situation. Photons with a few MeV energies in the comoving frame will
then be converted into pairs with very modest $\gamma$ provided that
the compactness parameter is more than unity. The pairs will generally
have relativistic energies, and will themselves participate in the
synchrotron and inverse Compton emission (Pe'er \& Waxman 2004).  In
reality the actual time dependence for an unsteady outflow leading to
shocks, pair formation, and Comptonisation could be more complicated.

If both a photospheric and a shock component are detected, one would
expect the thermal photospheric luminosity (and its non thermal part,
if present) to vary on similar timescales as the non thermal
synchrotron component (unless the shock efficiency is radius
dependent, or unless one or both are beyond $r = r_0 \eta^2$, in which
case $r/(c\eta^2)$ imposes a lower limit on the corresponding
variability timescale). The luminosity in a given band (e.g. {\it
Swift}) probably varies differently, since the thermal peak energy is
$\propto L^{1/4}$ and falls off steeply, while the synchrotron peak
energy varies $L^{3/2}$ and falls off more slowly. A preferred
low-energy break (Fig. 3) could be attributed to a scattered
photospheric component, provided the baryon loads are low or the
outflow variability timescales are large (Fig. 1). This still requires
a relatively strong shock synchrotron component, or possibly Alfv\'en
wave Comptonisation in the photosphere (e.g. Thompson 1994), to
explain the high-energy power-law spectra. It would also imply a
pronounced upward change of slope above the X-ray excess, from the
thermal peak to a flatter power law in all bursts where a low break is
observed. Further data on X-ray and $\gamma$-ray spectral features
during the burst (as opposed to the afterglow), will surely offer
important clues to the nature of the bulk flow and the macroscopic
source of energy driving the microscopic processes of particle
acceleration and cooling.

\section*{Acknowledgments}
I gratefully acknowledge valuable conversations with P. M\'esz\'aros,
A. Pe'er, M. J.~Rees and E. Waxman. I also thank the referee for
helpful correspondence. This work is supported by a Chandra
Postdoctoral Fellowship award PF3-40028.

\end{document}